\documentclass[a4paper]{jpconf}
\usepackage{graphicx}
\begin{document}
\title{Parity-projected Skyrme Hartree-Fock and angular momentum
projection approach to Mg isotopes
}

\author{Hirofumi Ohta$^{1,}$\footnote[3]{Present address:
Research Facility Center for Science and Technology, University of
 Tsukuba, Tsukuba 305-8577, Japan}, Takashi Nakatsukasa$^{1,2}$, and
 Kazuhiro Yabana$^{1,2}$} 

\address{$^1$Institute of Physics, University of Tsukuba, Tsukuba
 305-8571, Japan}
\address{$^2$Center for Computational Sciences, University of Tsukuba,
 Tsukuba 305-8571, Japan}

\ead{ohta@nucl.ph.tsukuba.ac.jp}

\begin{abstract}
We perform calculations of the variation after parity projection
 with Skyrme interaction for ground and excited states of even-even Mg
 isotopes.
Using the 3D real-space representation, we can take into account any
 kind of deformation; e.g., cluster structure, $\gamma$-deformation.
We also perform the full 3D angular momentum projection to obtained
 rotational spectra and $B(E2)$.
In $^{24}$Mg, the ground $K^\pi=0^+$ and several odd-parity bands are
 calculated.
In $^{30,32,34}$Mg, our calculation reproduces a breaking of the $N=20$
 magic number, so-called {\it Island of inversion} phenomena.
Properties of both the ground and excited states are reasonably
 reproduced.
\end{abstract}.

Among theories beyond the mean field, the variation after projection
 (VAP) method is one of the simplest ones.
Especially, it is known that the variation after parity projection
 (VAPP) is useful to describe exotic deformed solutions, octupole
 deformation or cluster structure within a restricted model space.
In addition, one can obtain negative-parity excited states which are
 either collective or non-collective modes of excitation.

Recently, we have proposed an algorithm to calculate the VAPP with the
 Skyrme interaction on the three-dimensional (3D) Cartesian coordinates
 (Parity Projected Skyrme-HF method = PPSHF) \cite{OYN04}.
In the ordinary mean-field calculations, one obtains self-consistent
 solutions with axial and reflection symmetries for most cases, so that
 the assumption of such spatial symmetry is reasonable to describe
 ground states.
However, to describe excited states, one has to take into account
 exotic shapes violating those symmetries.
In the VAPP calculations, self-consistent solutions often have
 symmetry-violating shapes because of the correlation beyond the
 mean-field.
In Ref.~\cite{OYN04}, we have shown that the cluster structure of
 $\alpha+^{16}$O appears in $^{20}$Ne and the three $\alpha$ structure
 in $^{12}$C, as a result of the VAPP in the 3D mesh representation.
% in $^{12}$C, as a result of the VAPP on 3D mesh representation using
% the Skyrme interaction.
We also perform the angular momentum projection (AMP) with 3D
 rotations.
Low-lying excitation spectra and transition strengths are well
 reproduced.

\begin{figure}[h]
\begin{minipage}{18pc}
\includegraphics[width=18pc]{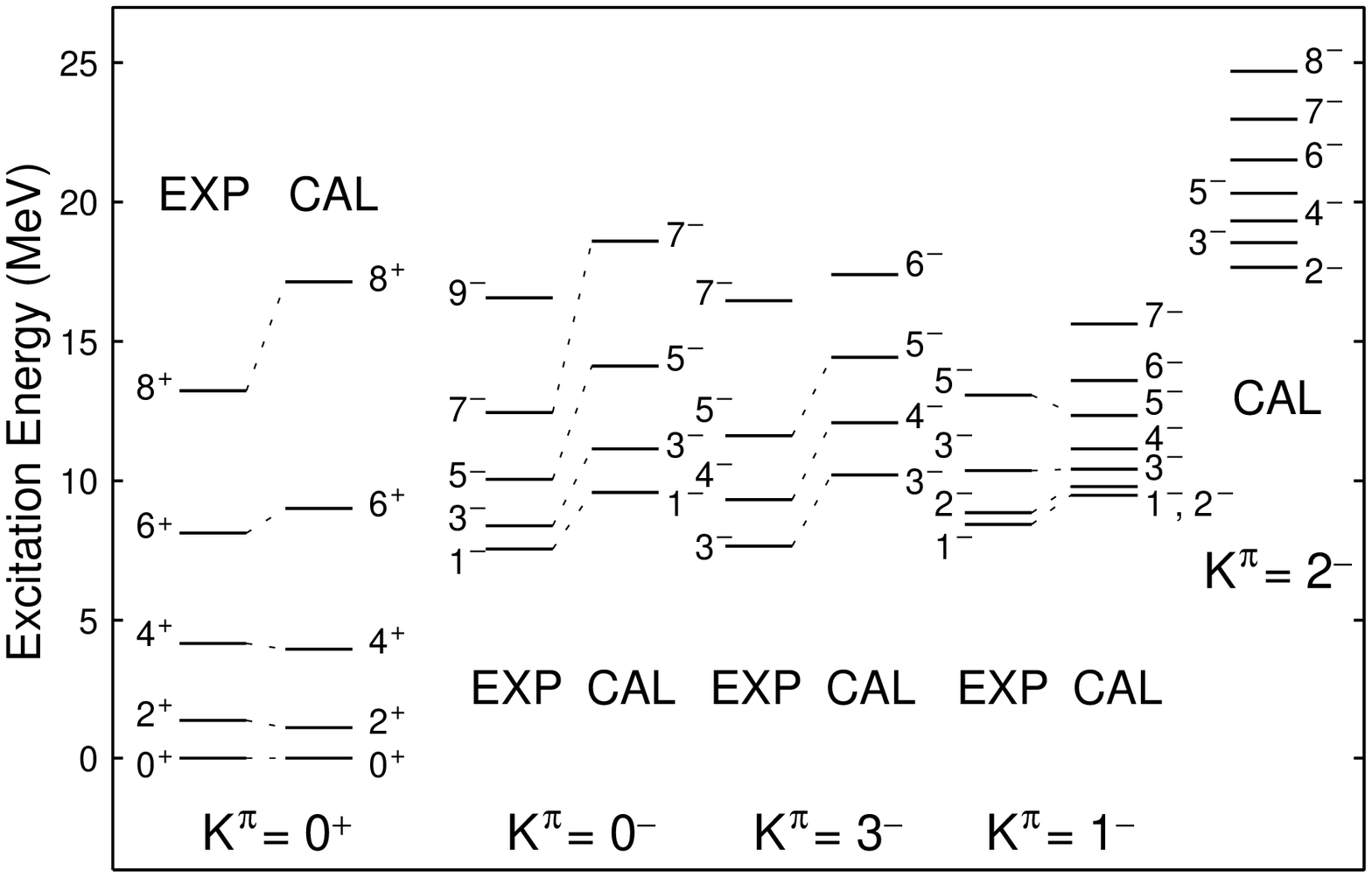}
\caption{\label{Mg24SPEC}The energy spectra for $^{24}$Mg calculated
 with PPSHF+AMP.
Experimental data are taken from \cite{FHIKKSU80}.}
\end{minipage}\hspace{2pc}
\begin{minipage}{18pc}
\includegraphics[width=18pc]{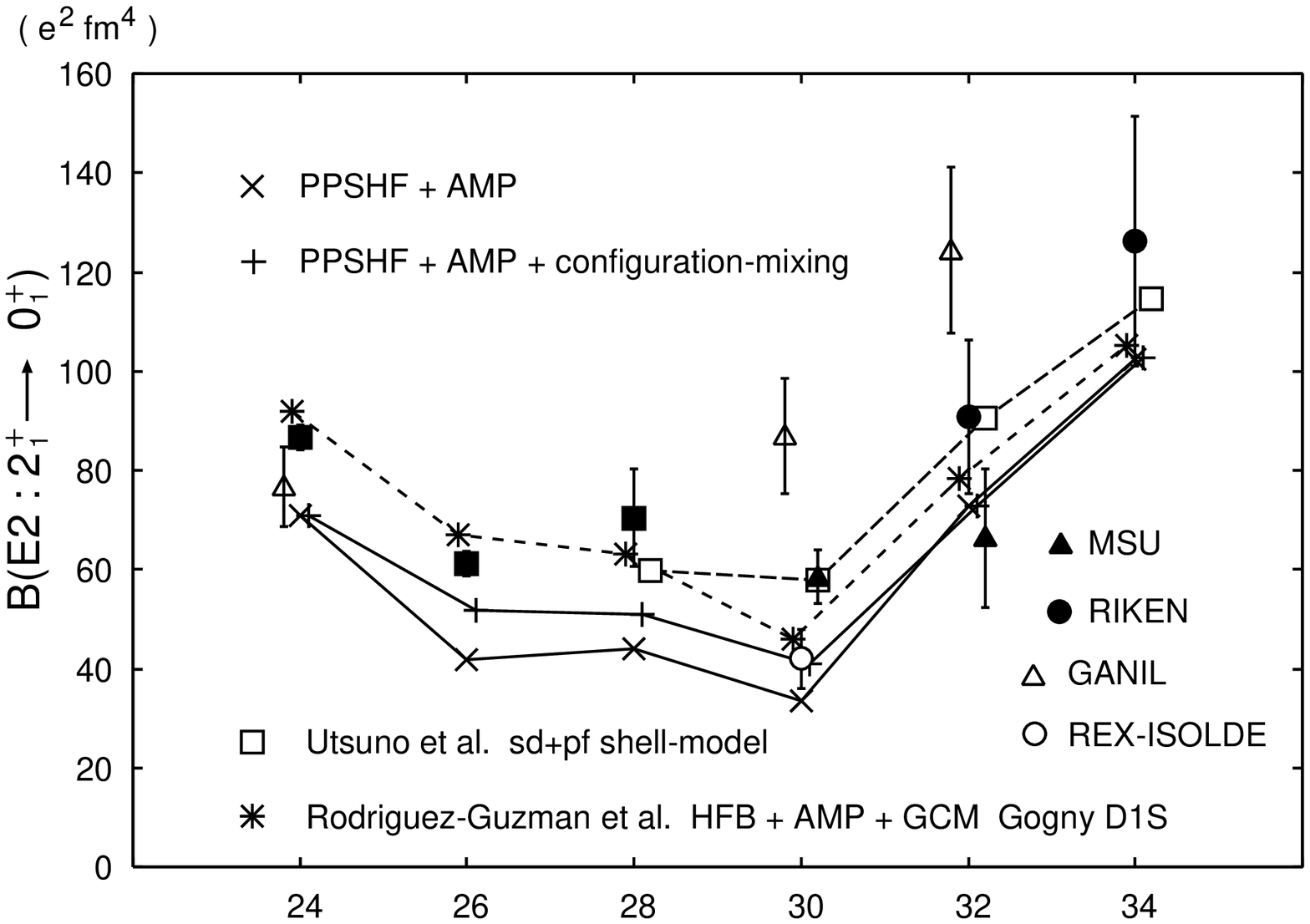}
\caption{\label{MgBE2}$B(E2:2^+_1\rightarrow0^+_1)$ of Mg isotopes.
Experimental results are from Coulomb excitation at
 RIKEN\cite{Motobayashi95,Iwasaki01}, MSU\cite{Pritychenko99} and
 REX-ISOLDE\cite{Niedermaier04}, inelastic scattering at
 GANIL\cite{Chiste01}. For comparison, we also show other calculations
 \cite{UOMH99,RER02}.}
\end{minipage} 
\end{figure}

In this study, we apply the method to Mg isotopes, $A=24$ to $34$.
The Skyrme functional with the SGII parameter set is used in the
 calculation.
We show the results of stable nucleus, $^{24}$Mg, in
 Fig.\ref{Mg24SPEC}.
We obtain a deformed ground state
%For the stable nucleus, $^{24}$Mg, we obtain a deformed ground state
 with positive parity, whose deformation is $\beta_2=0.52$ with some
 octupole components, $\beta_{30}=0.17$ and $\beta_{31}=0.12$.
The state has a dominant $K^\pi=0^+$ character.
The AMP reproduces experimental spectra and $B(E2)$ for the ground band
 up to $J^\pi=6^+$.
%The angular momentum projection reproduces experimental spectra and
% $B(E2)$ for the ground band up to $J^\pi=6^+$.
In addition, we also obtain negative-parity bands ($K^\pi=0^-,3^-,1^-$,
 and $2^-$), which well correspond to experimental data, though the
 calculated band-head energies are slightly higher than the experiments
 by $1-2.5$ MeV.

Next, let us discuss results for neutron-rich even-even Mg isotopes
 around the shell closure $N=20$, $^{30,32,34}$Mg.
In this region, the observed low excitation energy of the $2^+_1$ state
 and the enhanced $B(E2:0^+_1\rightarrow2^+_1)$ for $^{32}$Mg suggests
 an anomalous deformation and a quenching of the shell gap at $N=20$
 \cite{Motobayashi95,Iwasaki01,Pritychenko99}.
Despite the fact that the mean-field calculations with the Skyrme
 interaction have so far failed to reproduce these properties, our
 method reasonably reproduces observed data.
We show that the AMP is essential to reproduce
 these properties.
%We show that the angular momentum projection is essential to reproduce
% these properties.
In these nuclei, there are two positive-parity solutions, one of which
 has a large quadrupole deformation of $\beta\approx 0.4-0.6$ and the
 other has a small deformation of $\beta\approx 0.1-0.3$.
Calculated energies of these two states are close, and the interplay
 between these two seems to be a characteristic feature in the
 neutron-rich Mg isotopes.
In $^{30}$Mg, the one with a small deformation is the ground state,
 while the one with a large deformation becomes the lowest in
 $^{32,34}$Mg.
The ground state in $^{30}$Mg has a deformation of $\beta_2=0.22$ and
 $\beta_{33}=0.14$.
In $^{32}$Mg, we have the ground state with $\beta_2=0.44$ and
 $\beta_{33}=0.12$.
The rotational correlation provided by the AMP is essential to obtain
 the well-deformed ground state in $^{32}$Mg.
The deformation in $^{34}$Mg is even larger, $\beta_2=0.53$ with
 $\beta_{30}=0.17$.
This change of the ground-state character accounts for the experimental
 observation.
The calculated $B(E2;0^+\rightarrow 2^+)$ are also consistent with the
 experiments (Fig.\ref{MgBE2}).
Moments of inertia are somewhat overestimated in $^{30,32}$Mg, probably
 because the pairing correlation is neglected in the calculation.

%This work is supported by the Grant-in-Aid for Scientific Research in Japan
%(Nos. 14540369 and 14740146).

\smallskip

\end{document}